\documentclass[sigconf,nonacm]{acmart}
\usepackage{booktabs,array,graphicx,url}
\newcommand{\pc}[1]{#1\%}
\newcommand{\setupN}{2,660}
\newcommand{\setupAnyCount}{475}
\newcommand{\setupAnyPct}{17.9}
\newcommand{\setupSecurityCount}{409}
\newcommand{\setupSecurityPct}{15.4}
\newcommand{\setupConfigCount}{22}
\newcommand{\setupConfigPct}{0.8}
\newcommand{\setupSpecCount}{63}
\newcommand{\setupSpecPct}{2.4}
\newcommand{\setupPermissionCount}{165}
\newcommand{\setupPermissionPct}{6.2}

\newcommand{\setupRawRevisedCount}{488}
\newcommand{\setupRawRevisedPct}{18.3}
\newcommand{\collectionN}{511}
\newcommand{\collectionAnyCount}{35}
\newcommand{\collectionAnyPct}{6.8}

\newcommand{\setupMcpCount}{260}
\newcommand{\setupMcpPct}{9.8}
\newcommand{\setupMcpEligible}{1,063}
\newcommand{\setupMcpConditional}{24.5}

\newcommand{\setupGrantCount}{67}
\newcommand{\setupGrantPct}{2.5}
\newcommand{\setupGrantEligible}{533}
\newcommand{\setupGrantConditional}{12.6}

\newcommand{\setupSkillCount}{101}
\newcommand{\setupSkillPct}{3.8}
\newcommand{\setupSkillEligible}{1,988}
\newcommand{\setupSkillConditional}{5.1}
\newcommand{\collectionSkillCount}{19}
\newcommand{\collectionSkillPct}{3.7}

\newcommand{\setupAgentCount}{22}
\newcommand{\setupAgentPct}{0.8}
\newcommand{\setupAgentEligible}{773}
\newcommand{\setupAgentConditional}{2.8}

\newcommand{\setupNoFrontCount}{60}

\newcommand{\collectionNoFrontCount}{18}

\newcommand{\setupNoDescCount}{4}

\renewcommand\footnotetextcopyrightpermission[1]{}
\acmConference[]{}{}{}
\acmYear{2026}
\copyrightyear{2026}
\setcopyright{none}
\acmDOI{}
\acmISBN{}
\begin{document}
\title{Scanning the Harness: Configuration Exposures in AI Coding-Agent Supply Chains}
\author{Benjamin Kapner\textsuperscript{1,2}\quad Carmel Soceanu\textsuperscript{1}\quad Alicia Petrunin\textsuperscript{1}\quad Hofni Gartner\textsuperscript{1}}
\affiliation{\institution{\textsuperscript{1}Red Hat \qquad \textsuperscript{2}Stein Faculty of Computer and Information Science, Ben-Gurion University of the Negev}\country{}}
\renewcommand{\shortauthors}{Kapner et al.}

\begin{abstract}
AI coding agents rely on repository instructions, skills, hooks, tool-server
declarations, and subagent definitions. These artifacts distribute both
behavior and access to executable dependencies, making configuration review
part of the agent software supply chain. We study 3,171 public GitHub
repositories: 2,660 assembled setups and 511 skill collections. Deterministic
analysis, mechanical re-derivation, model-assisted adjudication, human
review, and platform documentation checks identify six categories of configuration exposure and
conformance issues. Unpinned MCP package declarations occur in
\pc{\setupMcpPct} of setups, broad execution grants in
\pc{\setupGrantPct}, and broad skill tool preapproval in
\pc{\setupSkillPct}. Their union covers \setupSecurityCount{} setups
(\pc{\setupSecurityPct}); among setups with MCP configuration,
\pc{\setupMcpConditional} contain an unpinned declaration. Including
required-field and skill-format issues brings the setup rate to
\pc{\setupAnyPct} and the collection rate to \pc{\collectionAnyPct}.
Holding the six categories fixed, contextual review changes the setup rate
from \pc{\setupRawRevisedPct} to \pc{\setupAnyPct}; rule selection explains
most of the reduction from the broader candidate set. The findings identify
concrete opportunities to pin dependencies, review execution preapproval,
and check component conformance before distribution or use. A documented
permission exception also exposes a shared error in the scanner and its
mechanical audit, motivating version-specific semantic checks. The study
provides reproducible evidence about repository declarations; agreement with human judgments informs label validation, while runtime
consequences and recall remain unmeasured.
\end{abstract}

\maketitle
\raggedbottom

\section{Introduction}
AI coding agents increasingly obtain their behavior from artifacts outside
the model: repository instructions, reusable skills, lifecycle hooks,
commands, MCP server declarations, and subagent definitions. We call this
configuration layer the agent's \emph{harness}. Its contents determine
which instructions are supplied, which external tools can be started, and
which permissions are requested or preapproved. Shared conventions such as
\texttt{AGENTS.md} make parts of this layer portable across
clients~\cite{agentsmd}.

Harness artifacts are distributed through repositories and marketplaces.
A skill can bundle instructions and scripts; a server declaration can
select an executable package; a permission entry can preapprove a broad
class of commands. These choices make configuration an actionable part of
the software supply chain. Published studies examine vulnerability
patterns in skill marketplaces and agent susceptibility to injected skill
instructions~\cite{liu2026wild,liu2026malicious,schmotz2026skillinject}.
We investigate how dependency declarations, execution permissions, and
component-format issues occur across both assembled setups and reusable
skill collections.

We analyze a released snapshot of 3,171 repositories with
\texttt{harness-\allowbreak eval}. A separate implementation checks the
scanner's predicates at pinned repository commits; model-assisted adjudication
examines disagreements and contextual cases; human review assesses the
repository-rule judgments; and documentation review checks platform behavior. The resulting six
categories cover three security exposures, one required agent field, and
two skill-specification issues.

The central result is that \setupSecurityCount{} of \setupN{} setups
(\pc{\setupSecurityPct}) contain an unpinned MCP package declaration or a
broad execution-preapproval declaration. Among setups with MCP
configuration, the unpinned-declaration rate is
\pc{\setupMcpConditional}. These findings locate decisions that maintainers
can inspect before adopting or sharing a harness: which package version
will be selected, and how much execution capability is preapproved.
Including required-field and skill-format issues raises the setup rate to
\pc{\setupAnyPct}. The unit of measurement is the repository declaration;
effective permissions and runtime consequences depend on the client and
deployment.

Reliable measurement also requires accurate client semantics. During
review, we found that wildcard \texttt{find} grants had been classified as
arbitrary execution because the utility supports \texttt{-exec}. Claude
Code explicitly excludes that form from wildcard approval
rules~\cite{claudepermissions}. Correcting this shared scanner and audit
error removes 62 findings. Separately, accounting for a fixed rule set
shows that category selection explains most of the difference between raw
and reported repository rates. These results make the interpretation of a
warning, as well as its detection, part of the measurement problem.

The paper contributes (1) a reproducible measurement of six configuration
categories across assembled setups and skill collections; (2) an accounting
that separates category selection, contextual review, and semantic
correction; and (3) practical guidance on dependency pinning, permission
review, and component conformance. All six categories are detectable at
file scope. Cross-file and graph alerts provide complementary evidence
about where contextual or runtime analysis is still needed.

\section{Background and Related Work}
\textbf{Harness components.} Context files supply persistent instructions.
Examples include \texttt{CLAUDE.md}, \texttt{AGENTS.md}, and
assistant-specific rules. Skills supply task-specific instructions and may bundle
scripts. Commands are user-triggered workflows. Hooks execute code at
lifecycle events. Subagent files describe delegated agents. MCP
declarations configure external tool providers. Loading, inheritance, and
permission semantics vary by client; recognizing a file format does not
establish its behavior in every client.

\textbf{Skill and agent security.} Liu et al.~\cite{liu2026wild} analyze
31,132 skills using static checks and model-based classification, with a
200-skill manually annotated validation set and reported precision and
recall. Their subsequent work examines malicious
skills~\cite{liu2026malicious}. Skill-Inject~\cite{schmotz2026skillinject}
is an attack benchmark measuring agent responses to injected skills, not a
marketplace prevalence study. Work on cross-tool attacks, taint-style
analysis, and MCP red-teaming studies runtime composition
risks~\cite{li2026dissonances,liu2025taint,he2026redteam}. Indirect prompt
injection and AgentDojo establish related attack and evaluation
settings~\cite{greshake2023injection,debenedetti2024agentdojo}.
Our study combines repository-level measurement, mechanical
re-derivation of the audited findings, and author-conducted manual review
of repository-rule classifications. These checks complement runtime
evaluation of agent behavior.

\textbf{Configuration and harness evaluation.} Infrastructure-as-code
analysis and configuration-smell studies establish a precedent for
structural checks on declarative
artifacts~\cite{chiari2022iac,sharma2016smell,rahman2019sins}. Studies of
agent context files examine their contents, smells, and effect on coding
performance~\cite{dossantos2026smells,chatlatanagulchai2025readmes,
gloaguen2026evaluating,lulla2026efficiency}. AutoSaddler optimizes harnesses
using execution traces~\cite{park2026autosaddler}, while ACES evaluates
skills through paired execution~\cite{kevin2026aces}. We address the
different question of configuration exposure and conformance across
component types.

\textbf{Analyzer evaluation.} Prior work documents the difficulty of
turning warnings into useful developer findings~\cite{johnson2013why,
bessey2010billion,kremenek2004ranking,austin2011comparison}. Our
predicate/consequence distinction applies that concern to agent
configuration. Mechanical agreement measures consistency between
implementations; human--pipeline agreement compares the recorded review judgments
with the pipeline labels.
Neither is, by itself, a precision estimate against independent ground
truth.

\section{Study Design}
\label{sec:design}
We ask: \textbf{RQ1}, how often do selected configuration conditions occur
in the corpus? \textbf{RQ2}, how do rates change with repository stratum,
component availability, and review policy? \textbf{RQ3}, which claims can
static analysis support, and which need additional runtime or contextual
evidence?

\subsection{Corpus and Units of Analysis}
The discovery pipeline uses assistant-specific GitHub topics, README
references to configuration filenames, thirty public curated lists, and
plugin marketplaces. These channels favor repositories that advertise or
distribute agent tooling. Of 9,295 candidates, the recorded funnel removes
3,065 with no detected component, 1,633 containing only instruction files,
1,322 with a single component type and fewer than five skills, 65 duplicates
or derived copies, and 22 clone/scan failures. A subsequent pass removes 17
more, leaving 3,171. These exclusion counts come from the released funnel
record; the artifact contains full scan results for the retained corpus,
rather than every excluded candidate.

The strata are mutually exclusive. A \emph{collection} has at least five
skills and no non-instruction component other than skills; a contributor
context file is allowed. This classification takes precedence. A
\emph{setup} otherwise has at least two detected component types. The
result is \setupN{} setups and \collectionN{} collections. Instruction-only
and sparse repositories are outside this operational population, although
some measured rules could apply to them. The two groups are useful views
of distributed artifacts and assembled configurations, but are not matched
suppliers and consumers.

The primary unit is a repository at a pinned commit. It contributes at
most once to each rule and once to a union, regardless of its number of
findings. The published deduplication uses commit identity and heuristics
based on repository name, component inventory, and descriptions; it does
not establish independence among copied skills or configurations.

\subsection{Instrument and Analysis Scope}
\texttt{harness-\allowbreak eval} discovers and parses component formats from several
agent clients and emits deterministic findings with rule, file, severity,
and remediation. The released records identify version 7.15.0 and two
scanner commits: 2,861 records name \texttt{eda472a}, and 310 name
\texttt{fad4a74}. Their recorded scan timestamps are September 6, 2026.
We reanalyze these archived records and preserve both scanner identifiers
so that each result remains traceable to its generating version.

The original analysis considers 31 candidate rules in four families:
security exposure (S), configuration integrity (Q), specification
conformance (P), and cross-assistant consistency (C). Twenty read one file
(FILE), seven also resolve filesystem context (FILE\_FS), and four compare
components (PAIRWISE). An exploratory graph analysis (SETUP), discussed
separately, searches explicit invocation edges between components. This
taxonomy describes the inputs the implemented check reads, not proof that
it completely captures the underlying security property.

\subsection{Exposure Definitions and Evidence}
\label{sec:construct}
A \emph{security exposure} here is a recorded declaration that leaves a
dependency version unspecified or requests broad execution preapproval.
The threat scenario is that a changed dependency or untrusted instruction
can exercise capability available to the agent. We do not assume that a
broad grant is accidental, that every component is loaded, or that the
declaration overrides the surrounding security policy. A \emph{configuration
issue} identifies an omitted field required for the intended component;
a \emph{specification issue} identifies a selected format deviation even
when a client tolerates it.

We distinguish three layers of evidence: the declaration exists; a
documented client gives that declaration a particular meaning; and an
actual deployment exercises it. The scan and re-derivation address the
first layer. Documentation review informs the second. This study does not
measure the third. A \emph{reported finding} is a declaration or field
omission included by the counting policy below.\footnote{We reserve
\emph{retained} for pair-level inclusion decisions in the audit.}

\subsection{Mechanical, Model-Assisted, and Human Review}
\label{sec:validation}
\textbf{Mechanical re-derivation.} A separate script re-clones flagged
repositories at their pinned commits and implements the documented
predicates. The full released audit contains 8,549 records over 1,115 corpus
repositories: 8,435 marked re-derived, 112 refuted, and two unverifiable.
The number of determinate outcomes is therefore 8,547. The audit also covers advisory checks
outside the final candidate set. Both implementations were developed by
the study team; agreement can expose implementation discrepancies but
cannot exclude shared errors.

\textbf{Contextual adjudication.} The original pipeline groups findings
into (repository, rule) pairs. A pair is agreed when all its findings are
re-derived and at least one is outside the predefined exclusions. A pair
is disputed if re-derivation fails, all findings belong to an excluded
subclass, or its consequence requires intent or external context. Examples
include fixtures, shipped templates, locally resolved packages, generated
files, and differences between assistants. A model receives the rule
definition and extracted evidence and returns \emph{defect},
\emph{not\_defect}, or \emph{uncertain}; only the first retains a disputed
pair. These are historical label names, not claims of ground truth.

The historical table contains 744 pairs across the primary, provisional,
and observation categories: 586 agreed and 158 adjudicated, with 640
pairs included by the original analysis. The saved adjudications identify the model as
\texttt{claude-fable-5-1}; the current script's default model is different,
so reproductions must use the recorded metadata rather than assume that
default describes the original run. No new model verdicts were generated
for this revision.

\textbf{Author-conducted manual review.} One author manually assessed
the repository-rule findings and recorded 752 human verdicts. Joining this
review to the final 744-pair historical table
by repository and rule yields 743 matched pairs. Agreement between human
verdicts and the historical pipeline labels is 93.3\% ($\kappa=0.76$);
for the 157 matched pairs requiring model adjudication, it is 70.7\%
($\kappa=0.23$). This manual review provides a human check of the
pipeline's classifications, distinct from the automated adjudication
described above. The comparison uses the final historical labels.

\textbf{Rule selection.} The original primary tier, called ``gating'' in
the artifact, requires at least 50 determinate findings, at least 97\%
mechanical agreement, and at least 80\% retained repository-rule pairs.
Six rules meet those thresholds. A provisional rule has 13--49
determinate findings and complete mechanical agreement; it is excluded
from headline unions. Six mechanically consistent rules fail the
consequence threshold and are reported as contextual observations. The
remaining candidates have inadequate counts, disagreement, or no findings.
These are descriptive selection rules, not calibrated deployment guarantees
or a preregistered test of a detector.

\subsection{Semantic Correction and Reanalysis}
\label{sec:correction}
We interpret declarations against documentation accessed September 24,
2026. The original execution categories included wildcard
\texttt{find} grants because the program can execute another command.
However, Claude Code documents that these wildcard grants do not
auto-approve its \texttt{-exec} and \texttt{-delete} forms. We therefore
exclude that pattern in both settings and skill preapproval
categories~\cite{claudepermissions}. This removes 36 settings findings and
26 skill findings, eliminating 16 settings pairs and one skill pair. A
repository with another eligible finding remains counted. The remaining
grant patterns are interpreted as broad capability declarations under the
evidence model in Section~\ref{sec:construct}.

All other historical inclusion decisions remain fixed. The reanalysis
script verifies the artifact commit, reproduces the original headline
counts, applies the exclusion, and writes revised counts, input hashes,
excluded findings, and retained pairs. Table~\ref{tab:audit} reports the
remaining findings. The exclusion ledger makes the semantic correction reproducible at both
finding and repository-rule level.

We report exact counts and proportions of this corpus. We omit population
confidence intervals because discovery is not probability sampling and
repositories can share components. For context, we also report proportions
among setups in which the scanner detects the relevant component type.
Those denominators describe opportunities for the check to apply, not
observed runtime exposure.

\section{Results}
\label{sec:results}
\subsection{Corpus Composition and Findings}
The median setup contains six components and 7,057 estimated tokens; the
largest recorded values are 500 components and 1,611,649 tokens. The
scanner's median estimate of always-loaded content is 1,198 tokens for
setups and 189 for collections. These are inventory estimates, not
measurements of actual model context after client loading and truncation.
Among setups, \setupSkillEligible{} contain skills,
\setupMcpEligible{} MCP configuration, \setupAgentEligible{} subagents,
and \setupGrantEligible{} hook/settings components.

\begin{figure*}[t]
\centering
\includegraphics[width=\textwidth]{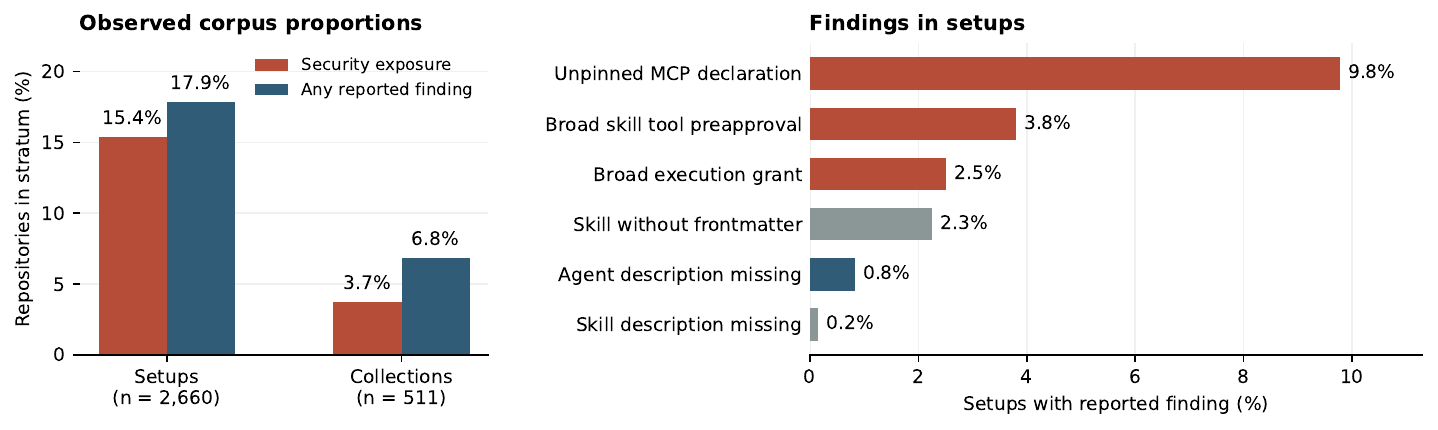}
\Description{Two bar charts show observed proportions of repositories with
reported findings. Setups have 15.4 percent security exposure and 17.9
percent any reported finding; collections have 3.7 and 6.8 percent,
respectively. The second chart separates the six setup categories.}
\caption{Observed proportions in the sampled repositories. The left panel
shows the union of security exposures and the union of all six categories.
The right panel separates the setup categories; repositories can contribute
to more than one category.}
\label{fig:results}
\end{figure*}

Table~\ref{tab:results} and Figure~\ref{fig:results} summarize the six
reported categories. The three security-exposure categories affect
\setupSecurityCount{} setups (\pc{\setupSecurityPct}). Missing agent
descriptions affect \setupConfigCount{} (\pc{\setupConfigPct}); selected
skill-specification deviations affect \setupSpecCount{}
(\pc{\setupSpecPct}). Their union is \setupAnyCount{}
(\pc{\setupAnyPct}). Categories overlap, so row counts cannot be added.
Collections contain \collectionAnyCount{} repositories with a reported
finding (\pc{\collectionAnyPct}).

\begin{table}[t]
\caption{Repository counts and within-stratum proportions after the semantic
correction. S: security exposure; Q: configuration integrity; P:
specification conformance. Denominators are 2,660 setups and 511 collections.
Zero counts can reflect absent component types.}
\label{tab:results}
\small\centering
\begin{tabular}{@{}p{.48\columnwidth}rr@{}}
\toprule
Finding & Setups & Collections \\
\midrule
Unpinned MCP declaration (S) & 260 (9.8\%) & 0 (0.0\%) \\
Broad execution grant (S) & 67 (2.5\%) & 0 (0.0\%) \\
Broad skill tool preapproval (S) & 101 (3.8\%) & 19 (3.7\%) \\
Agent description missing (Q) & 22 (0.8\%) & 0 (0.0\%) \\
Skill without frontmatter (P) & 60 (2.3\%) & 18 (3.5\%) \\
Skill description missing (P) & 4 (0.2\%) & 0 (0.0\%) \\
\midrule
Any security exposure (S) & 409 (15.4\%) & 19 (3.7\%) \\
Any configuration issue (Q) & 22 (0.8\%) & 0 (0.0\%) \\
Any specification issue (P) & 63 (2.4\%) & 18 (3.5\%) \\
Any reported finding & 475 (17.9\%) & 35 (6.8\%) \\
\bottomrule
\end{tabular}

\end{table}

\subsection{Dependency and Permission Declarations}
\label{sec:security}
\textbf{Unpinned MCP packages.} \setupMcpCount{} setups
(\pc{\setupMcpPct}) contain an unpinned package declaration, typically a
runner such as \texttt{npx -y @scope/server} without a version. Among the
\setupMcpEligible{} setups with MCP configuration, the proportion is
\pc{\setupMcpConditional}. The original review included 260 of 272
flagged repository-rule pairs; its exclusions include fixtures, templates,
and project-local dependencies.

The exposure is that dependency resolution is left unpinned: installations
or updates can select different executable versions from the same harness
declaration.\footnote{Caching affects when resolution occurs:
\texttt{npm exec} can use locally installed or cached packages, and Docker's
default pull policy reuses an available image~\cite{npmexec,dockerrun}.}
Pinning the package version or image digest makes that choice explicit,
while controlling transitive dependencies requires an appropriate lockfile
or resolved manifest.

\textbf{Broad execution grants.} \setupGrantCount{} setups
(\pc{\setupGrantPct}) contain settings entries for unrestricted
\texttt{Bash} or broad access to an execution-capable command. The category
covers shells, interpreters, package runners, and selected utilities such
as \texttt{awk} and \texttt{sed}. For example,
\texttt{Bash(python:*)} places no restriction
on which Python program is named in the grant. Such declarations warrant
review of the required automation, available credentials, and surrounding
policy. The conditional proportion is \pc{\setupGrantConditional} among
\setupGrantEligible{} setups with hook/settings components, an inventory
denominator broader than setups with an explicit allow list.

\textbf{Skill tool preapproval.} \setupSkillCount{} setups
(\pc{\setupSkillPct}) and \collectionSkillCount{} collections
(\pc{\collectionSkillPct}) contain broad \texttt{allowed-tools} entries.
Among setups with skills, the proportion is \pc{\setupSkillConditional}.
Claude Code applies this preapproval during the invoking turn and clears
it at the next user message. The field adds preapproval rather than
restricting the available tool set; ask/deny rules continue to
apply~\cite{claudeskills}. Other clients can interpret or support the field
differently~\cite{skillspec}.

Settings and skill preapproval together occur in
\setupPermissionCount{} setups (\pc{\setupPermissionPct}). Their security
relevance is the breadth of execution capability they request. Maintainers
can compare that breadth with the intended task and narrow grants where
more specific commands suffice.

\subsection{Configuration and Specification Issues}
\setupAgentCount{} setups lack a required description in a reported
subagent finding. This is \pc{\setupAgentPct} of all setups and
\pc{\setupAgentConditional} of the \setupAgentEligible{} with subagent
components. Claude Code documents skipping a named subagent without a
description; GitHub's custom-agent format also requires a
description~\cite{claudesubagents,copilotagents}. A required-field check
therefore offers an actionable diagnostic before an agent definition is
loaded.

The Agent Skills specification requires frontmatter with \texttt{name}
and \texttt{description}, whereas Claude Code accepts omitted fields and
provides fallbacks~\cite{skillspec,claudeskills}. Missing frontmatter occurs
in \setupNoFrontCount{} setups and \collectionNoFrontCount{} collections;
a present block without a description occurs in \setupNoDescCount{}
setups. The setup union is \setupSpecCount{} because one repository appears
in both categories. These selected deviations identify portability and
conformance issues even when a permissive client can load the skill.

\subsection{Separating Rule Selection from Review}
\label{sec:selection}
Comparing raw and reviewed rates requires a fixed set of rules. Raw output
across 31 candidate categories flags 678 setups (25.5\%). Restricting that
output to the six reported categories gives 503 (18.9\%); after correcting
the wildcard \texttt{find} interpretation, it gives
\setupRawRevisedCount{} (\pc{\setupRawRevisedPct}). Contextual inclusion
then yields \setupAnyCount{} (\pc{\setupAnyPct}).

Table~\ref{tab:funnel} separates the effects in an explicit order. Selecting
the six categories removes 175 repositories from the raw union; excluding
the \texttt{find} pattern removes another 15; applying the historical
contextual inclusion decisions removes another 13. Because conditions
overlap, these marginal reductions depend on the order and should not be
interpreted as independent causal effects. For collections, the same six
raw categories already yield 35 repositories and the subsequent steps do
not change that union.

\begin{table}[t]
\caption{Accounting for the setup union, with a fixed denominator of 2,660.
Each row applies the stated restriction to the preceding row.}
\label{tab:funnel}
\small\centering
\begin{tabular}{@{}p{.65\columnwidth}rr@{}}
\toprule
Analysis stage & Count & Share \\
\midrule
Raw output, 31 candidate categories & 678 & 25.5\% \\
Raw output, six selected categories & 503 & 18.9\% \\
Same six, excluding wildcard \texttt{find} & \setupRawRevisedCount & \setupRawRevisedPct\% \\
Same six, historical pair decisions applied & \setupAnyCount & \setupAnyPct\% \\
\bottomrule
\end{tabular}
\end{table}

A sensitivity check that rejects every historically disputed pair gives
474 setups (17.8\%) in the corrected six-rule union; accepting all gives
488 (18.3\%). This small band addresses only those recorded disagreements.
Shared semantic errors require a separate check: the \texttt{find}
correction affected pairs that the two implementations had agreed on.

\subsection{Comparing Repository Strata}
Collections lack MCP and hook/settings components by definition. Their
zero counts for server declarations and settings grants are consequently
structural zeros, not evidence that the supplying population is safer.
Skill conditions are observable in both strata. Broad skill preapproval
appears in \pc{\setupSkillPct} of all setups, but
\pc{\setupSkillConditional} of setups that actually contain skills,
compared with \pc{\collectionSkillPct} of collections. Changing the
denominator changes the comparison without changing a single finding.

The comparison supports two points of review. Publishers can inspect skill
fields before distribution; consumers also need to inspect the settings
and server declarations in their assembled configurations. Establishing
transfer between the two strata would require matched content or
installation histories, which this corpus does not provide.

\subsection{Contextual Observations and Graph Alerts}
\label{sec:observations}
Six mechanically consistent categories require contextual interpretation.
Table~\ref{tab:observations} reports their historical model-assisted
outcomes separately. Differences between instruction files can be
assistant-specific variants; missing paths can be generated or external;
an apparently incomplete MCP declaration can be a shipped template.
The contextual review included some alerts and excluded others, motivating
case-specific diagnostics for these categories.

\begin{table}[t]
\caption{Historical contextual categories, outside the six-rule union.
Included pairs are those accepted by the original model-assisted review.}
\label{tab:observations}
\small\centering
\begin{tabular}{@{}p{.61\columnwidth}rr@{}}
\toprule
Condition & Pairs & Included \\
\midrule
Missing context-file import & 23 & 11 \\
Agent references an absent skill & 7 & 3 \\
Context files differ across assistants & 71 & 29 \\
MCP declarations differ across assistants & 22 & 8 \\
MCP endpoint-integrity warning & 5 & 1 \\
MCP structural-validity warning & 13 & 1 \\
\bottomrule
\end{tabular}
\end{table}

The exploratory credential-to-network graph rule flagged six
repositories. The released reading describes one token-counting pattern
mistaken for credential access and five ordinary uses of a key with its
issuing service. None of the six established the claimed cross-component
exfiltration path. The result identifies a specific challenge for graph analysis: an
invocation edge must be connected to evidence of sensitive-data transfer
before it supports an exfiltration claim. Recall remains unmeasured.

An additional reference-resolution rule flags 33.1\% of setups and 64.0\%
of collections. An exploratory automated classification of 112 sampled
alerts assigns 49 to dead paths, 21 to a real file referenced by the wrong
path, and 42 to planned outputs. These exploratory labels suggest a useful distinction for future tooling: an input expected to exist and an
output that an agent is instructed to create should not be tested by the
same existence check.

\section{Implications}
\label{sec:implications}
\textbf{Configuration review has concrete targets.} The
\pc{\setupSecurityPct} result locates \setupSecurityCount{} setups with a
dependency-selection or execution-preapproval decision worth reviewing.
The \pc{\setupMcpConditional} rate among MCP-bearing setups makes version
pinning a particularly relevant check for server declarations. These
findings can be inspected before running an agent, and the six reported
categories need only file-level checks. Table~\ref{tab:recommendations}
assigns the corresponding actions to the parties able to take them.

\textbf{Prioritize by deployment context.} Under a threat model involving
untrusted instructions or changing dependencies, execution grants and
package resolution deserve attention before portability warnings. Broad
preapproval can increase the actions available without another prompt;
unpinned resolution can change the executable selected by a stable
configuration. The practical priority depends on activation, sandboxing,
credentials, and overriding policies. Missing agent descriptions deserve
attention when the affected agent is needed, while skill-format deviations
matter particularly for distribution across clients. This is a
threat-model-based review order; the study does not estimate incident
likelihood or quantify severity.

\begin{table*}[t]
\caption{Recommended actions and the parties responsible for them. These
recommendations follow from the measured declarations and documented
semantics; their effectiveness has not been experimentally evaluated.}
\label{tab:recommendations}
\small\centering
\begin{tabular}{@{}p{.18\textwidth}p{.24\textwidth}p{.53\textwidth}@{}}
\toprule
Responsible party & Review target & Recommended action \\
\midrule
Setup maintainers & MCP dependency declarations & Pin intended package versions or image digests; record resolved dependencies and review updates. \\
Setup maintainers and skill publishers & Broad execution preapproval & Check the commands needed by the task; narrow grants where feasible and document required capabilities. \\
Agent and skill publishers & Required fields and portable formats & Supply agent descriptions and specification-required skill metadata; check the formats of intended clients before release. \\
Agent-client developers & Dependency and permission review & Show resolved versions and the effective scope of grants, including the effect of trust, deny/ask rules, and sandboxing. \\
Analyzer authors and evaluators & Rule semantics and measurement & Version client-specific rules, test documented exceptions, and report fixed-rule counts alongside independent validation. \\
\bottomrule
\end{tabular}
\end{table*}

\textbf{Make client semantics part of the instrument.} The 62 excluded
\texttt{find} findings show how a second implementation can reproduce the
same mistaken model of a client. Versioned behavioral cases should cover
both broad grants and documented exceptions. Scope also matters: Claude
Code documents that project/local \texttt{bypass\allowbreak Permissions}
stopped taking effect in v2.1.257, while \texttt{dontAsk} denies actions
that would need approval~\cite{claudesettings,claudepermissions}. These
settings require distinct interpretations and are outside the revised
primary union. Analyzer reports should expose the matching declaration,
the applicable client semantics, and the evidence for any claimed
consequence.

\textbf{Separate contextual diagnostics from automatic fixes.} Cross-file
differences can express intentional client variants, and referenced paths
can name outputs an agent is expected to create. Input/output annotations
and client-specific validation could improve these checks. Additional
human assessment of unflagged cases and runtime evaluation are needed to
measure recall and the consequences of the reported exposures.

\section{Threats to Validity}
\label{sec:threats}
\textbf{Construct validity.} The study measures selected declarations and
field omissions. A security exposure is not an exploit, a broad capability
need not be unwanted, and a specification deviation need not cause a
failure. We have not measured users' expectations, deployment policies,
workspace trust, activation, or attacker control. Documented semantics can
differ by client version and operating system. Our correction addresses a
known \texttt{find} exception; the remaining permission patterns have not
been exhaustively tested. The rule vocabulary omits some code-executing
build tools and Git grants. Selected skill-field checks likewise do not
cover every specification requirement, and transitive dependency
resolution is outside the measurement.

\textbf{Label validity.} Shared rule definitions can induce correlated
errors in both implementations and model adjudication. One author performed
the manual review; familiarity with the instrument may introduce confirmation
bias. Human--pipeline agreement measures neither inter-human reliability
nor held-out detector accuracy or deployment consequences. The adjudicator's
judgment of whether a maintainer would fix a condition does not substitute
for maintainer feedback. The exploratory reference-resolution labels lack
separate human assessment. The disagreement sensitivity band covers
recorded disputes, not shared errors. Reviewing unflagged cases is necessary
to assess recall.

\textbf{Selection and external validity.} Discovery favors visible,
advertised configurations; the inclusion rules exclude instruction-only
and sparse repositories. Repository size increases the opportunity to
trigger at least one finding, and the reported proportions are not weighted
by users, downloads, skills, or installations. Heuristic deduplication does
not remove every shared template and can also merge distinct repositories.
Private enterprise configurations may differ. We therefore report corpus
proportions without ecosystem-wide sampling claims or confidence intervals.
Conditional component denominators improve interpretation but do not
control for these differences.

\textbf{Reproducibility and scope.} Input repository commits, scanner
commits, and output records are pinned; the snapshot contains two scanner
commits under one version string. Reanalysis can reproduce its counts but
is not evidence of an identical fresh scan. Historical label corrections
and rule selection were performed on the same corpus. No held-out detector
evaluation, runtime exploit study, or installation-provenance analysis was
conducted. False negatives and residual false positives are unknown, so
neither the primary proportions nor the zero confirmed graph alerts are
formal bounds on security defects.

\begin{table*}[t]
\caption{Mechanical agreement and included pairs after the semantic correction.
Re-derived is an implementation-agreement count. Included pairs apply the
historical inclusion policy plus the documented \texttt{find} exclusion.}
\label{tab:audit}
\small\centering
\begin{tabular}{@{}lrrrr@{}}
\toprule
Category & Audit findings & Re-derived & Flagged repository-rule pairs & Included pairs \\
\midrule
Unpinned MCP declaration & 961 & 961 & 272 & 260 \\
Broad execution grant & 125 & 125 & 67 & 67 \\
Broad skill tool preapproval & 872 & 872 & 120 & 120 \\
Agent description missing & 158 & 157 & 23 & 22 \\
Skill without frontmatter & 306 & 306 & 78 & 78 \\
Skill description missing & 63 & 63 & 4 & 4 \\
\bottomrule
\end{tabular}

\end{table*}

\section{Conclusion}
Agent configuration is an inspectable part of the software supply chain.
Across \setupN{} assembled setups, \setupSecurityCount{}
(\pc{\setupSecurityPct}) contain an unpinned MCP package declaration or
broad execution preapproval. Required agent fields and selected skill-format
issues bring the total to \setupAnyCount{} setups
(\pc{\setupAnyPct}); \collectionAnyCount{} of \collectionN{} skill
collections (\pc{\collectionAnyPct}) also contain a reported finding.
These results identify practical checks for dependency selection,
permission review, and component conformance at publication and adoption.

The study also shows why configuration analysis needs explicit semantic
validation. Correcting a documented permission exception changed counts
even where the scanner and mechanical audit agreed. Fixed-rule accounting,
versioned client semantics, and traceable evidence make these measurements
reproducible and their conclusions assessable.

\section*{Data Availability}
The scanner is available at
\url{https://github.com/redhat-community-ai-tools/harness-eval}.
The original corpus, audit, adjudications, and reader table are available at
\url{https://github.com/Benkapner/harness-eval-experiments}, commit
\texttt{217b8620e8a7} (full identifier in the supplement).
The accompanying \texttt{revision/} supplement provides reanalysis and
figure scripts, input hashes, revised counts, pair and exclusion tables,
reproduction instructions, and a proposed validation protocol.
Its \texttt{reader/} directory preserves all human verdicts and agreement
statistics, restores the original human-review column names, and records
the author's confirmation of their provenance. One credential-like literal
is redacted from scanner evidence. The public commit remains the immutable
input baseline.

\section*{Ethical Considerations}
The study uses public repository contents and released records; no
repository-provided code or exploit was executed for this revision.
Repository identities and commit links remain in the artifact. The revised
manuscript and supplement reproduce no credential values. Original
secret-related alerts require separate handling before redistribution,
irrespective of the graph-alert results. An author performed the manual
review; no external human reviewers were recruited.

\section*{Acknowledgment of AI Assistance}
OpenAI Codex~\cite{codex} assisted with manuscript-wide prose revision,
reanalysis and visualization code, and formatting. It generated no new
corpus observations or adjudication labels. Section~\ref{sec:validation}
distinguishes model adjudication from the author's manual review; writing
assistance did not generate the human verdicts.

\appendix
\section{Mechanical Agreement for the Six Categories}
\label{sec:appendix}
Table~\ref{tab:audit} separates finding-level re-derivation from
repository-level inclusion. The settings and skill categories exclude the
62 \texttt{find} findings. A positive pair can contain several findings,
and a high agreement rate does not establish runtime correctness. The
historical broader review counts in Section~\ref{sec:validation} refer to
the original table, not a new human or model review of the corrected categories.

\end{document}